\documentclass[10pt,a4paper,onecolumn]{article}
\usepackage{marginnote}
\usepackage{graphicx}
\usepackage{xcolor}
\usepackage{authblk,etoolbox}
\usepackage{titlesec}
\usepackage{calc}
\usepackage{tikz}
\usepackage{hyperref}
\hypersetup{colorlinks,breaklinks=true,
            urlcolor=[rgb]{0.0, 0.5, 1.0},
            linkcolor=[rgb]{0.0, 0.5, 1.0}}
\usepackage{caption}
\usepackage{tcolorbox}
\usepackage{amssymb,amsmath}
\usepackage{seqsplit}
\usepackage{xstring}
\usepackage{orcidlink}

\usepackage{float}
\let\origfigure\figure
\let\endorigfigure\endfigure
\renewenvironment{figure}[1][2] {
    \expandafter\origfigure\expandafter[H]
} {
    \endorigfigure
}

\usepackage[
  backend=biber,
]{biblatex}
\bibliography{paper.bib}

\let\textttOrig=\texttt
\def\texttt#1{\expandafter\textttOrig{\seqsplit{#1}}}
\renewcommand{\seqinsert}{\ifmmode
  \allowbreak
  \else\penalty6000\hspace{0pt plus 0.02em}\fi}

\makeatletter
\let\href@Orig=\href
\def\href@Urllike#1#2{\href@Orig{#1}{\begingroup
    \def\Url@String{#2}\Url@FormatString
    \endgroup}}
\def\href@Notdoi#1#2{\def\tempa{#1}\def\tempb{#2}%
  \ifx\tempa\tempb\relax\href@Urllike{#1}{#2}\else
  \href@Orig{#1}{#2}\fi}
\def\href#1#2{%
  \IfBeginWith{#1}{https://doi.org}%
  {\href@Urllike{#1}{#2}}{\href@Notdoi{#1}{#2}}}
\makeatother

\newlength{\cslhangindent}
\newlength{\csllabelwidth}
\newenvironment{CSLReferences}[3] 
 {
  \setlength{\parindent}{0pt}
  \ifodd #1 \everypar{\setlength{\hangindent}{\cslhangindent}}\ignorespaces\fi
  \ifnum #2 > 0
  \setlength{\parskip}{#2\baselineskip}
  \fi
 }%
 {}
\usepackage{calc}

\usepackage[top=3.5cm, bottom=3cm, right=1.5cm, left=1.0cm,
            headheight=2.2cm, reversemp, includemp, marginparwidth=4.5cm]{geometry}

\titleformat{\section}
  {\normalfont\sffamily\Large\bfseries}
  {}{0pt}{}
\titleformat{\subsection}
  {\normalfont\sffamily\large\bfseries}
  {}{0pt}{}
\titleformat{\subsubsection}
  {\normalfont\sffamily\bfseries}
  {}{0pt}{}
\titleformat*{\paragraph}
  {\sffamily\normalsize}

\usepackage{fancyhdr}
\makeatletter
\let\ps@plain\ps@fancy
\fancyheadoffset[L]{4.5cm}
\fancyfootoffset[L]{4.5cm}

\definecolor{linky}{rgb}{0.0, 0.5, 1.0}

\newtcolorbox{repobox}
   {colback=red, colframe=red!75!black,
     boxrule=0.5pt, arc=2pt, left=6pt, right=6pt, top=3pt, bottom=3pt}

\newcommand{\ExternalLink}{%
   \tikz[x=1.2ex, y=1.2ex, baseline=-0.05ex]{%
       \begin{scope}[x=1ex, y=1ex]
           \clip (-0.1,-0.1)
               --++ (-0, 1.2)
               --++ (0.6, 0)
               --++ (0, -0.6)
               --++ (0.6, 0)
               --++ (0, -1);
           \path[draw,
               line width = 0.5,
               rounded corners=0.5]
               (0,0) rectangle (1,1);
       \end{scope}
       \path[draw, line width = 0.5] (0.5, 0.5)
           -- (1, 1);
       \path[draw, line width = 0.5] (0.6, 1)
           -- (1, 1) -- (1, 0.6);
       }
   }

\patchcmd{\@maketitle}{center}{flushleft}{}{}
\patchcmd{\@maketitle}{center}{flushleft}{}{}
\patchcmd{\@maketitle}{\LARGE}{\LARGE\sffamily}{}{}
\def\maketitle{{%
  
  \AB@maketitle}}
\makeatletter
\renewcommand\AB@affilsepx{ \protect\Affilfont}
\renewcommand\AB@affilnote[1]{{\bfseries #1}\hspace{3pt}}
\renewcommand{\affil}[2][]%
   {\newaffiltrue\let\AB@blk@and\AB@pand
      \if\relax#1\relax\def\AB@note{\AB@thenote}\else\def\AB@note{#1}%
        \setcounter{Maxaffil}{0}\fi
        \begingroup
        \let\href=\href@Orig
        \let\texttt=\textttOrig
        \let\protect\@unexpandable@protect
        \def\thanks{\protect\thanks}\def\footnote{\protect\footnote}%
        \@temptokena=\expandafter{\AB@authors}%
        {\def\\{\protect\\\protect\Affilfont}\xdef\AB@temp{#2}}%
         \xdef\AB@authors{\the\@temptokena\AB@las\AB@au@str
         \protect\\[\affilsep]\protect\Affilfont\AB@temp}%
         \gdef\AB@las{}\gdef\AB@au@str{}%
        {\def\\{, \ignorespaces}\xdef\AB@temp{#2}}%
        \@temptokena=\expandafter{\AB@affillist}%
        \xdef\AB@affillist{\the\@temptokena \AB@affilsep
          \AB@affilnote{\AB@note}\protect\Affilfont\AB@temp}%
      \endgroup
       \let\AB@affilsep\AB@affilsepx
}
\makeatother

\renewcommand\Affilfont{\sffamily\small\mdseries}
\IfFileExists{upquote.sty}{\usepackage{upquote}}{}
\IfFileExists{microtype.sty}{%
\usepackage{microtype}
\UseMicrotypeSet[protrusion]{basicmath} 
}{}

\usepackage{hyperref}
\hypersetup{unicode=true,
            pdftitle={raccoon: A Python package for removing wiggle artifacts in the JWST NIRSpec integral field spectroscopy},
            pdfborder={0 0 0},
            breaklinks=true}
\let\addcontentslineOrig=\addcontentsline
\def\addcontentsline#1#2#3{\bgroup
  \let\texttt=\textttOrig\addcontentslineOrig{#1}{#2}{#3}\egroup}
\let\markbothOrig\markboth
\def\markboth#1#2{\bgroup
  \let\texttt=\textttOrig\markbothOrig{#1}{#2}\egroup}
\let\markrightOrig\markright
\def\markright#1{\bgroup
  \let\texttt=\textttOrig\markrightOrig{#1}\egroup}

\usepackage{graphicx,grffile}
\makeatletter
\def\maxwidth{\ifdim\Gin@nat@width>\linewidth\linewidth\else\Gin@nat@width\fi}
\def\maxheight{\ifdim\Gin@nat@height>\textheight\textheight\else\Gin@nat@height\fi}
\makeatother
\setkeys{Gin}{width=\maxwidth,height=\maxheight,keepaspectratio}
\IfFileExists{parskip.sty}{%
\usepackage{parskip}
}{
\setlength{\parindent}{0pt}
\setlength{\parskip}{6pt plus 2pt minus 1pt}
}
\ifx\paragraph\undefined\else
\let\oldparagraph\paragraph
\renewcommand{\paragraph}[1]{\oldparagraph{#1}\mbox{}}
\fi
\ifx\subparagraph\undefined\else
\let\oldsubparagraph\subparagraph
\renewcommand{\subparagraph}[1]{\oldsubparagraph{#1}\mbox{}}
\fi

\title{raccoon: A Python package for removing wiggle artifacts in the
JWST NIRSpec integral field spectroscopy}

        \author[1, 2, 3]{Anowar J. Shajib\orcidlink{0000-0002-5558-888X}}
    
      \affil[1]{Department of Astronomy \& Astrophysics, University of
Chicago, Chicago, IL 60637, USA}
      \affil[2]{Kavli Institute for Cosmological Physics, University of
Chicago, Chicago, IL 60637, USA}
      \affil[3]{Center for Astronomy, Space Science and Astrophysics,
Independent University, Bangladesh, Dhaka 1229, Bangladesh}
  \date{\vspace{-7ex}}

\begin{document}
\maketitle

\marginpar{

  \begin{flushleft}
  \sffamily\small

  {\bfseries DOI:} \href{https://doi.org/10.xxxx}{\color{linky}{10.xxxx}}

  \vspace{2mm}

  {\bfseries Software}
  \begin{itemize}
    \setlength\itemsep{0em}
    \item \href{N/A}{\color{linky}{Review}} \ExternalLink
    \item \href{https://github.com/ajshajib/raccoon}{\color{linky}{Repository}} \ExternalLink
    \item \href{DOI unavailable}{\color{linky}{Archive}} \ExternalLink
  \end{itemize}

  \vspace{2mm}

  \par\noindent\hrulefill\par

  \vspace{2mm}

  {\bfseries Editor:} \href{https://example.com}{Pending
Editor} \ExternalLink \\
  \vspace{1mm}
    {\bfseries Reviewers:}
  \begin{itemize}
  \setlength\itemsep{0em}
    \item \href{https://github.com/Pending Reviewers}{@Pending
Reviewers}
    \end{itemize}
    \vspace{2mm}

  {\bfseries Submitted:} N/A\\
  {\bfseries Published:} N/A

  \vspace{2mm}
  {\bfseries License}\\
  Authors of papers retain copyright and release the work under a Creative Commons Attribution 4.0 International License (\href{http://creativecommons.org/licenses/by/4.0/}{\color{linky}{CC BY 4.0}}).

  \end{flushleft}
}

\hypertarget{summary}{%
\section{Summary}\label{summary}}

\texttt{raccoon} is a Python package for removing resampling noise --
commonly referred to as ``wiggles'' -- from spaxel-level spectra in
datacubes obtained from the JWST Near Infrared Spectrograph's (NIRSpec)
integral field spectroscopy (IFS) mode. These wiggles arise as artifacts
during resampling of the 2D raw data into 3D datacubes, due to the point
spread function (PSF) being undersampled. The standard JWST data
reduction pipeline does not correct for this noise. The wiggle artifacts
can significantly degrade the scientific usability of the data,
particularly at the spaxel level, undermining the exquisite spatial
resolution of NIRSpec. \texttt{raccoon} provides an empirical correction
by modeling and removing these artifacts, thereby restoring the fidelity
of the extracted spectra. \texttt{raccoon} forward-models the wiggles as
a chirp function impacting one or more template spectra that are
directly fit to the original data across the entire wavelength range.
The best-fit wiggle model is then used to clean the data while
propagating the associated uncertainties.

\hypertarget{statement-of-need}{%
\section{Statement of need}\label{statement-of-need}}

The JWST NIRSpec's IFS mode (\hyperlink{ref-Boker23}{Böker et al., 2023}) enables spatially
resolved infrared spectroscopy of astronomical sources with an
unprecedented combination of signal-to-noise ratio, redshift coverage,
and spatial resolution. However, it suffers from heavy undersampling of
the point-spread function, leading to resampling noise that manifests as
low-frequency sinusoidal artifacts known as ``wiggles.'' These artifacts
can significantly distort the overall spectral shape, bias line
measurements, and compromise kinematic analyses at the single-spaxel
level, thereby limiting the scientific potential of the NIRSpec IFU
data. \texttt{raccoon} provides a user-friendly, robust, and
computationally efficient solution to identify and remove these
artifacts, enabling precise studies of galaxy kinematics and
early-Universe phenomena that would otherwise be hindered by the
presence of wiggles. \texttt{raccoon} has already been used in
scientific publications to facilitate robust measurements of stellar
kinematics from JWST NIRSpec spectra (\hyperlink{ref-TDCOSMO25}{TDCOSMO
Collaboration, 2025}; \hyperlink{ref-Shajib25}{Shajib et al., 2025}).

\hypertarget{functionality}{%
\section{Functionality}\label{functionality}}

\begin{figure}
\centering
\includegraphics{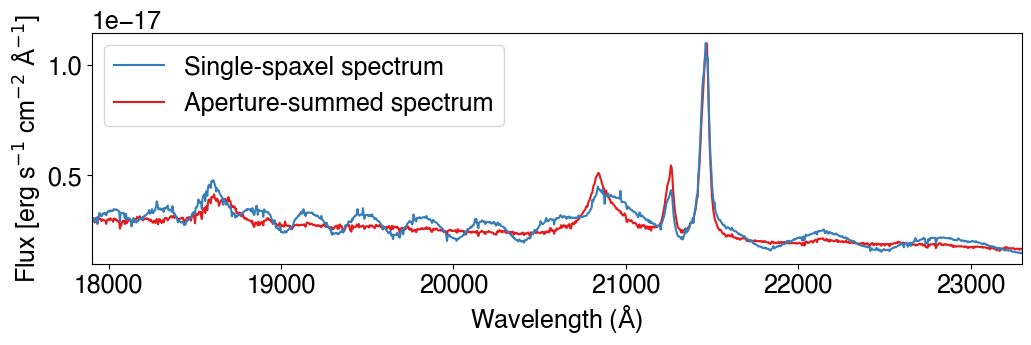}
\caption{\label{fig:wiggled-in-spectra}
Illustration of the wiggles in the single-spaxel spectrum
(blue), which is a manifestation of the resampling noise in the
standard-pipeline-reduced datacube due to PSF undersampling. The red
spectrum shows the aperture-summed spectrum within a 4-spaxel radius,
normalized to match its maximum to that of the single-spaxel one. The
illustrated data are of an active galactic nucleus (AGN) from Perna et al. (2023).}
\end{figure}

Since the wiggles in the single-spaxel spectra in the reduced JWST
NIRSpec datacube are artifacts or resampling noise arising due to
undersampling the PSF, dithering can, in principle, mitigate the issue
of PSF undersampling and thus the wiggles. However, the commonly adopted
4-point dither pattern is still insufficient to recover the optimal
sampling and completely eliminate the wiggles in the rectified datacube
(\hyperlink{ref-Law23}{Law et al., 2023}). The wiggles are washed out in the spectra summed
from multiple spaxels within a sufficiently large aperture (illustrated
in \autoref{fig:wiggled-in-spectra}). This provides a basis for making
an empirical correction for the wiggles by comparing the single-spaxel
spectrum to the one summed within an aperture around it. This principle
was also used by previous corrective algorithms, such as in \hyperlink{ref-Perna23}{Perna et al.
(2023)} and in the Python routine \texttt{WICKED} (\hyperlink{ref-Dumont25}{Dumont et al., 2025}).

In \texttt{raccoon}, the wiggle is modeled as a sinusoidal chirp
function \[ 
    W(\lambda) = 1 + A(\lambda) \left[ \sin \phi(\lambda) + a_1 \sin^2 \phi (\lambda) + a_2 \sin(3\phi (\lambda)) \right], \tag{1}\label{eq:wiggle} 
\] where \(A(\lambda)\) is the wavelength-dependent amplitude and
\(\phi (\lambda) = \lambda\,k(\lambda) + \phi_0\) is the phase that
depends on a wavelength-dependent wavenumber \(k(\lambda)\). The
wiggle's peaks and troughs can be asymmetrically and symmetrically
sharpened (or de-sharpened) by varying the \(a_1\) and \(a_2\)
parameters, respectively.

Given this parametrization for the wiggle, a single-spaxel spectrum
\(D(\lambda)\) is modeled with \[ 
    M (\lambda) = W(\lambda) \, T(\lambda), \tag{2}\label{eq:model} 
\] where \(T(\lambda)\) is a template for the correct spectrum devoid of
the wiggles. In \texttt{raccoon}, this template is constructed based on
the circular-aperture-summed spectrum \(C(\lambda)\). The template can
also optionally include a spectrum \(S(\lambda)\) summed from a shell or
annulus centered on the corresponding spaxel, following \hyperlink{ref-Dumont25}{Dumont et al.
(2025)}. Including the shell-summed spectrum in the template can account
for changes in the line shape between the single-spaxel spectrum and the
aperture-summed one, for example, in the case of lines broadened by
stellar kinematics that can vary across the galaxy (\hyperlink{ref-Dumont25}{Dumont et al.,
2025}). The aperture radius and the inner and outer radii of the shell
are to be adjusted by the user, as the appropriate values can depend on
the source morphology and the astronomical scene. The template is
constructed as \[ 
    T(\lambda) = c_1\,C(\lambda) + c_2 \, S(\lambda) + c_3 \,\lambda^b + \sum_{n=0}^N p_n\, \lambda^n ,  \tag{3}\label{eq:template} 
\] where \(c_1,c_2, c_3, p_0, \dots, p_N\) are linear coefficients.
Here, the power-law plus polynomial term (i.e.,
\(c_3 \,\lambda^b + \sum_{n=0}^N p_n\, \lambda^n\)) on the right-hand
side models the difference in the continuum between the single-spaxel
spectrum and \(c_1\,C(\lambda) + c_2 \, S(\lambda)\).

\begin{figure}
\hypertarget{fig:full-fit-example}{%
\centering
\includegraphics{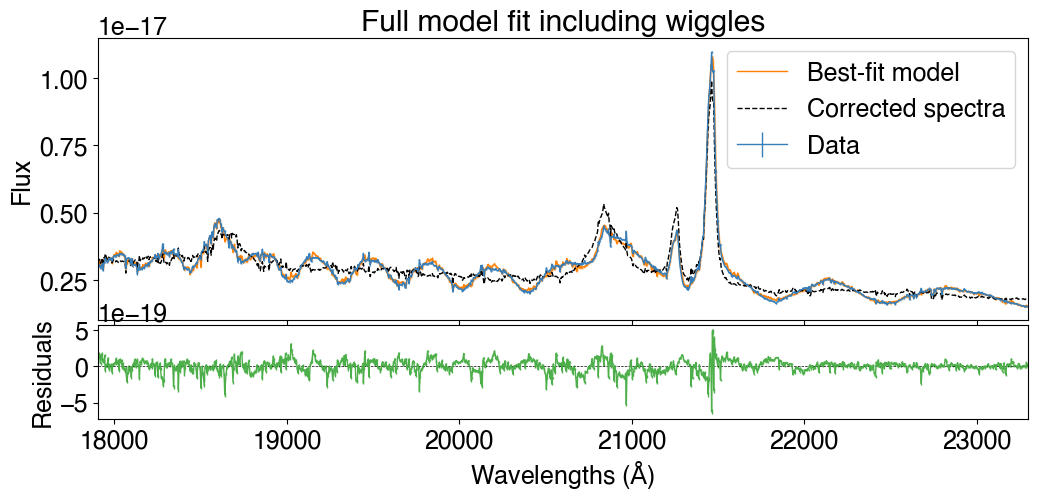}
\caption{Modeling of the full spectrum (blue in the top panel) based on
the template spectra and the wiggles impacting it. The illustrated
spectrum is the same one from \autoref{fig:wiggled-in-spectra}. The
best-fit model is shown in orange and the wiggle-corrected spectrum is
shown in black. The bottom panel illustrates the residuals (green)
between the original data and the best-fit
model.}\label{fig:full-fit-example}
}
\end{figure}

\begin{figure}
\hypertarget{fig:wiggle-model-example}{%
\centering
\includegraphics{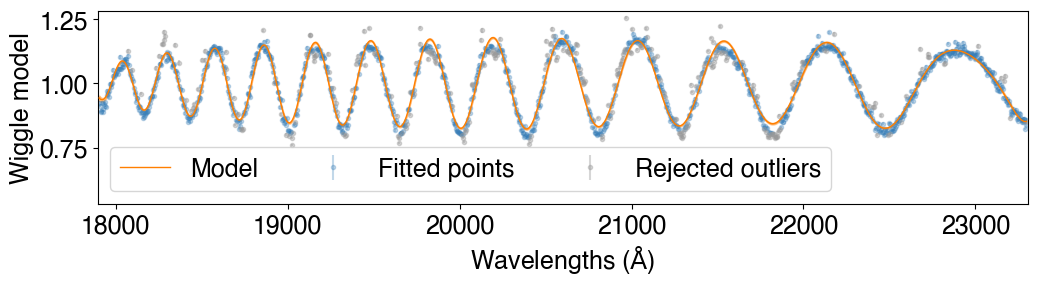}
\caption{Illustration of the extracted and modeled wiggle signal. The
illustrated data points represent \(D(\lambda)/M(\lambda)\) based on the
best-fit \(M(\lambda)\), and the orange line illustrates the best-fit
wiggle model \(W(\lambda)\). The grey points are rejected outliers using
the false discovery rate method, and the blue points mark the
wavelengths where the data were fit to the
model.}\label{fig:wiggle-model-example}
}
\end{figure}

The functions \(A(\lambda)\) and \(k(\lambda)\) are modeled with
B-splines, with the number of knots adjustable by the user.
\texttt{raccoon} provides functionality for the user to determine the
appropriate number of knots using model selection criteria based on the
Bayesian information criterion (BIC) or the minimum \emph{a posteriori}
chi-square metric (\(\chi^2_{\rm MAP}\)). The best-fit values for the
linear coefficients (i.e., \(c_1, c_2, c_3, p_0, \dots, p_N\)) and
non-linear parameters (i.e., \(a_1\), \(a_2\), \(b\), and the
coefficients of the B-spline basis functions) are determined by
minimizing \[ 
    \chi^2 = \sum_{i} \frac{(D_i - M_i)^2}{\sigma_i^2}, \tag{4}\label{eq:chisquare} 
\] where the index \(i\) runs across the wavelength pixels and
\(\sigma_{i}\) is the associated noise level.
\autoref{fig:full-fit-example} illustrates an example of the fitted
model to a given spectrum (of an AGN). In this example,
\texttt{raccoon}'s robust performance is demonstrated in fitting the
given spectrum while modeling the wiggle signal
(\autoref{fig:wiggle-model-example}). The user can mask out regions of
the spectrum that have strong features potentially impacting the quality
of the fit, or optionally adopt outlier rejection in the fit using the
false discovery rate method (\hyperlink{ref-Benjamini95}{Benjamini \& Hochberg, 1995}) or
sigma-clipping.

\texttt{raccoon} enables users to efficiently process multiple spaxels
in a datacube, either across the entire field or within a specified
spatial region. A user-defined detection threshold for the wiggle signal
can be set, ensuring that corrections are only applied to spectra where
significant wiggles are present.

\texttt{raccoon} offers several advantages over previously available
scripts and routines. Earlier tools (\hyperlink{ref-Perna23}{Perna et al.,
2023}; \hyperlink{ref-Dumont25}{Dumont et al., 2025}) extract the wiggle signal by subtracting the single-spaxel
spectrum from a template spectrum (similar to or the same as that in
\autoref{eq:template}) fitted to the data without accounting for
wiggles. This approach can result in the loss of a fraction of the
wiggle signal during numerical fitting. In contrast, \texttt{raccoon}
fits the template spectrum and the parametric wiggle model
\(W(\lambda)\) simultaneously to the data, eliminating the need for an
intermediate extraction step and preserving the full wiggle signal.
While previous scripts perform fitting separately in multiple wavelength
slices, \texttt{raccoon} fits across the entire wavelength range at
once. This makes \texttt{raccoon} more robust to gaps in the fitted
spectrum, whether excluded by manual masking or outlier rejection.
Another important distinction is that earlier tools apply an additive
correction, whereas \texttt{raccoon} uses a multiplicative correction.
Since both approaches derive the correction empirically from the data,
this difference should not affect performance in practice. However, as
the wiggles manifest as a multiplicative effect on the spectra (\hyperlink{ref-Law23}{Law et
al., 2023}), \texttt{raccoon}'s implementation is physically motivated.
Additionally, \texttt{raccoon} is uniquely installable via the
\texttt{pip} command, allowing more flexible import into Python scripts
or notebooks and making it more user-friendly and portable than its
peers.

\hypertarget{acknowledgements}{%
\section{Acknowledgements}\label{acknowledgements}}

The author acknowledges helpful discussions with Michele Cappellari,
Frédéric Courbin, David Law, and Tommaso Treu. AJS received support from
NASA through STScI grants JWST-GO-2974 and HST-GO-16773. This work makes
use of the \texttt{Astropy}, \texttt{NumPy}, \texttt{SciPy}, and
\texttt{Matplotlib} packages.

\hypertarget{references}{%
\section*{References}\label{references}}
\addcontentsline{toc}{section}{References}

\hypertarget{refs}{}
\begin{CSLReferences}{1}{0}
\leavevmode\hypertarget{ref-Benjamini95}{}%
Benjamini, Y., \& Hochberg, Y. (1995). Controlling the false discovery
rate: A practical and powerful approach to multiple testing.
\emph{Journal of the Royal Statistical Society: Series B
(Methodological)}, \emph{57}(1), 289--300.
\url{https://doi.org/10.1111/j.2517-6161.1995.tb02031.x}

\leavevmode\hypertarget{ref-Boker23}{}%
Böker, T., Beck, T. L., Birkmann, S. M., Giardino, G., Keyes, C.,
Kumari, N., Muzerolle, J., Rawle, T., Zeidler, P., Abul-Huda, Y., Alves
de Oliveira, C., Arribas, S., Bechtold, K., Bhatawdekar, R.,
Bonaventura, N., Bunker, A. J., Cameron, A. J., Carniani, S., Charlot,
S., \ldots{} Willott, C. J. (2023). In-orbit {Performance} of the
{Near-infrared Spectrograph NIRSpec} on the {James Webb Space
Telescope}. \emph{Publications of the Astronomical Society of the
Pacific}, \emph{135}, 038001.
\url{https://doi.org/10.1088/1538-3873/acb846}

\leavevmode\hypertarget{ref-Dumont25}{}%
Dumont, A., Neumayer, N., Seth, A. C., Böker, T., Eracleous, M., Goold,
K., Greene, J. E., Gültekin, K., Ho, L. C., Walsh, J. L., \&
Lützgendorf, N. (2025). {WIggle Corrector Kit for NIRSpEc Data: WICKED}.
\emph{arXiv e-Prints}, arXiv:2503.09697.
\url{https://doi.org/10.48550/arXiv.2503.09697}

\leavevmode\hypertarget{ref-Law23}{}%
Law, D. R., E. Morrison, J., Argyriou, I., Patapis, P., Álvarez-Márquez,
J., Labiano, A., \& Vandenbussche, B. (2023). A {3D Drizzle Algorithm}
for {JWST} and {Practical Application} to the {MIRI Medium Resolution
Spectrometer}. \emph{The Astronomical Journal}, \emph{166}, 45.
\url{https://doi.org/10.3847/1538-3881/acdddc}

\leavevmode\hypertarget{ref-Perna23}{}%
Perna, M., Arribas, S., Marshall, M., D'Eugenio, F., Übler, H., Bunker,
A., Charlot, S., Carniani, S., Jakobsen, P., Maiolino, R., Rodríguez Del
Pino, B., Willott, C. J., Böker, T., Circosta, C., Cresci, G., Curti,
M., Husemann, B., Kumari, N., Lamperti, I., \ldots{} Scholtz, J. (2023).
{GA-NIFS}: {The} ultra-dense, interacting environment of a dual {AGN} at
z {\(\sim\)} 3.3 revealed by {JWST}/{NIRSpec IFS}. \emph{679}, A89.
\url{https://doi.org/10.1051/0004-6361/202346649}

\leavevmode\hypertarget{ref-Shajib25}{}%
Shajib, A. J., Treu, T., Suyu, S. H., Law, D., Yıldırım, A., Cappellari,
M., Galan, A., Knabel, S., Wang, H., Birrer, S., Courbin, F., Fassnacht,
C. D., Frieman, J. A., Melo, A., Morishita, T., Mozumdar, P., Sluse, D.,
\& Stiavelli, M. (2025). {TDCOSMO XXIII. First spatially resolved
kinematics of the lens galaxy obtained using JWST-NIRSpec to improve
time-delay cosmography}. \emph{arXiv e-Prints}, arXiv:2506.21665.
\url{http://arxiv.org/abs/2506.21665}

\leavevmode\hypertarget{ref-TDCOSMO25}{}%
TDCOSMO Collaboration. (2025). {TDCOSMO 2025: Cosmological constraints
from strong lensing time delays}. \emph{arXiv e-Prints},
arXiv:2506.03023. \url{https://doi.org/10.48550/arXiv.2506.03023}

\end{CSLReferences}

\end{document}